# A Comparison of Deep Learning Convolution Neural Networks for Liver Segmentation in Radial Turbo Spin Echo Images

Lavanya Umapathy[1,2], Mahesh Bharath Keerthivasan[1,2], Jean-Phillipe Galons[2], Wyatt Unger[2], Diego Martin[2], Maria I Altbach[2], and Ali Bilgin[1,2,3]

[1]Department of Electrical and Computer Engineering, University of Arizona, Tucson, Arizona
[2]Department of Medical Imaging, University of Arizona, Tucson, Arizona
[3]Department of Biomedical Engineering, University of Arizona, Tucson, Arizona

## ABSTRACT
Motion-robust 2D Radial Turbo Spin Echo (RADTSE) pulse sequence can provide a high-resolution composite image, T2-weighted images at multiple echo times (TEs), and a quantitative T2 map, all from a single k-space acquisition. In this work, we use a deep-learning convolutional neural network (CNN) for the segmentation of liver in abdominal RADTSE images. A modified UNET architecture with generalized dice loss objective function was implemented. Three 2D CNNs were trained, one for each image type obtained from the RADTSE sequence. On evaluating the performance of the CNNs on the validation set, we found that CNNs trained on TE images or the T2 maps had higher average dice scores than the composite images. This, in turn, implies that the information regarding T2 variation in tissues aids in improving the segmentation performance.

## INTRODUCTION
T2-weighted imaging is widely used for diagnosis in hepatic applications and there is a growing role for quantitative imaging in the evaluation of hepatic diseases[1]. Radial Turbo Spin Echo[2] (RADTSE) sequence has been used in abdominal imaging since it is motion-insensitive and produces high-resolution images of the liver within a breath-hold. Figure 1 shows an illustration of the processing of k-space data acquired using RADTSE sequence. A single RADTSE k-space data set can be used to obtain a high-resolution composite image, T2-weighted images at multiple echo times (TEs) using echo sharing[3], and a T2 map for quantitative evaluation. Importantly, all of these different images are co-registered.

The segmentation of liver is the first step in the automated processing of abdominal RADTSE images. Since all images (composite, multiple TEs, and T2 map) are co-registered, segmentation masks for liver obtained using one image type can be easily applied to the others. In this work, we investigate the use of deep-learning Convolutional Neural Networks (CNN) for the segmentation of RADTSE images. We evaluate the performance of deep-learning CNN segmentation trained using different types of images obtained from RADTSE data to determine the best image type to use for segmentation.

## MATERIALS AND METHODS
### Convolutional Neural Network
2D CNNs with U-NET[4] like architecture have been successful in various medical image segmentation tasks. In this work, we modify the 2D UNET[4] architecture for the segmentation of RADTSE images (Figure 2). Here, a batch normalization layer is used after every convolution layer; the convolutions are zero padded to retain the input image size at the output. The number of output features at convolutional layer are depicted in Figure 2.

### Imaging Data
A cohort of 32 subjects were imaged on a Siemen 1.5T (Erlangen, Germany) scanner with the following imaging protocol: image matrix: 320x320, in-plane resolution: 1.56mm x 1.56mm, slice thickness: 8 mm, TR: 2500ms, echo spacing: 7.3ms, echo train length: 32. Of these, 29 subjects were used for training and 3 were retained for testing. The ground truth annotations for liver were manually drawn by an expert radiologist. The pre-processing of these images included N4 bias correction algorithm[5], followed by contrast stretching to enhance the contrast between the liver and the background tissues. The composite and TE images were normalized using zero mean and unit standard deviation normalization. The T2 maps were clipped to limit the range of intensities to [0, 500] ms. Image augmentation included random translations, blurring using Gaussian filters, and elastic deformations.

**Network Implementation**

In order to compare the segmentation performance with the different images, we implemented three different CNNs. The CNNs to be trained on composite images and T2 maps were implemented to accept a single-channel (SC-UNET) input whereas the CNN to be trained in multiple TE images was implemented to accept multi-channel (MC-UNET) input. Two separate SC-UNETs were trained; one for the composite images (SC-UNET-Composite) and the other for T2 maps (SC-UNET-T2). A two-channel MC-UNET was trained using T2-weighted images acquired at TE=14ms and TE=100ms. These TE images were concatenated along the channel dimension. Figure 3 compares the contrast between composite, multiple TE images, and the T2 map for a representative imaging slice.

A generalized dice loss[6] (GDL) [6] with an $L_2$ penalty was implemented as the loss function. This loss function is defined as follows:

$$GDL = 1 - \frac{\sum_n p_n r_n}{\sum_n p_n + \sum_n r_n} - \frac{\sum_n (1-p_n)(1-r_n)}{\sum_n p_n + \sum_n r_n}$$

Here, $r_n$ and $p_n$ refer to the $n^{th}$ pixel in the ground truth annotation and prediction, respectively. The loss function gradients were calculated and implemented to work with the Caffe[7] framework.

A coarse search for the best learning-rate (LR) identified a LR of 0.001 with consistent performance for all the three nets. Each net was randomly initialized and trained with the following parameters: Iterations = 75000 iterations, optimizer = stochastic gradient descent, learning rate = 0.001, momentum = 0.99, decay = 0.1. All networks were trained on a PC with multiple NVIDIA Titan X GPUs. The generalizing ability of the CNNs was evaluated using dice overlap metric.

**RESULTS**

The end to end prediction time for a single image was approximately 80ms on average for the three CNNs.

Figure 4 shows representative images from three subjects, ground truth annotations for liver, and the predicted liver masks for the three CNNs.

The mean and the standard deviation for dice for the three CNNs are shown in Table 1. It can be observed that both MC-UNET and SC-UNET-T2 show improved dice scored compared to the SC-UNET-Composite. However, SC-UNET-T2 has a lower standard deviation value implying consistency in performance. The improved performances of SC-UNET-T2 and MC-UNET can be attributed to the fact that the input images to these networks contain information regarding quantitative T2 variations at each pixel.

**CONCLUSION**

Three different convolutional nets were implemented to compare liver segmentation performance on images obtained from the RADTSE sequence. We observed that networks using multiple TE images (MC-UNET) or T2 maps (SC-UNET-T2) as input yielded better dice scores and sensitivity compared to the one using only composite images as inputs (SC-UNET-Composite).

**ACKNOWLEDGEMENTS**

The authors would like to acknowledge support from the Arizona Biomedical Research

Commission (Grant ADHS14-082996) and the Technology and Research Initiative Fund (TRIF) Improving Health Initiative

# FIGURES

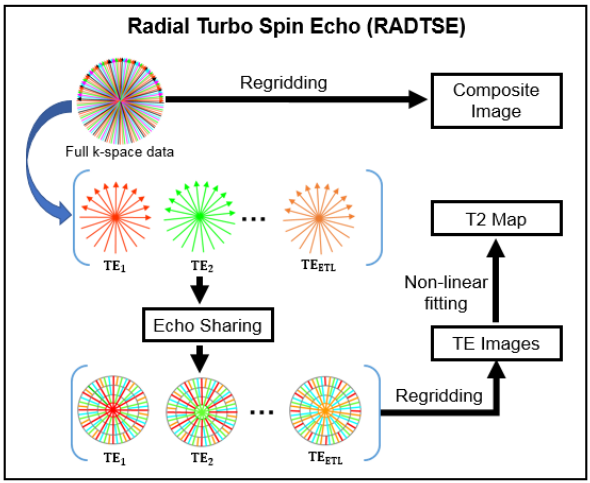

**Figure 1.** Illustration of the processing of Radial Turbo Spin Echo (RADTSE) data. The full k-space radial data can be regridded to obtain a composite image. Echo sharing of the undersampled k-space data from individual TEs can generate TE images using echo sharing techniques. A non-linear fitting of the TE images can generate T2 maps. Note that the composite images, TE images, and the T2 maps are obtained from the same k-space data, and are co-registered.

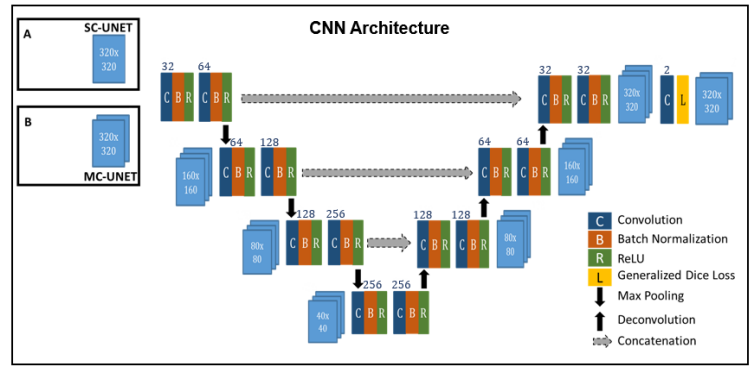

**Figure 2.** Illustration of the modified UNET CNN used in this study. This architecture has been modified to include four resolution levels, batch normalization layer after every convolution layer, zero padded convolutions, and a generalized dice loss layer. Based on the input, the network can either be single channel (SC-UNET) or multi-channel (MC-UNET).

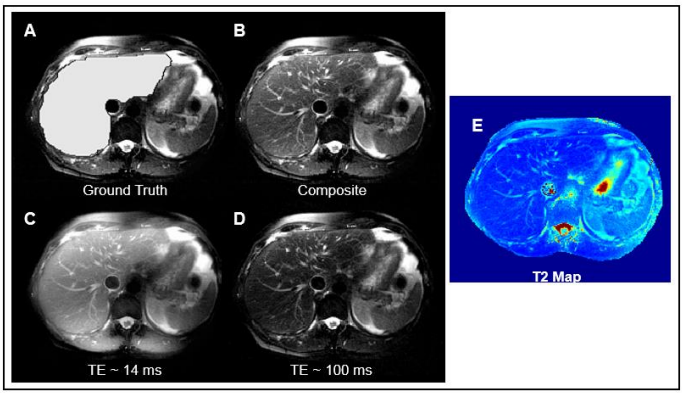

**Figure 3.** Representative images from three subjects with the ground truth mask overlaid on images are shown in (A). The predictions for liver masks for the three CNNS (B, C and D) are overlaid on the respective input images.

|  | Mean | Std. Dev |
|---|---|---|
| SC-UNET-Composite | 0.822 | 0.45 |
| MC-UNET | 0.883 | 0.41 |
| SC-UNET-T2 | 0.889 | 0.18 |

**Table 1.** A comparison of the dice overlap scores (mean and standard deviation) between the different CNNs on the test subjects are shown here. Although the CNNs trained on multiple TE images (MC-UNET) and T2 map (SC-UNET-T2) have mean dice scores that are comparable, the standard deviation bounds are tighter for SC-UNET-T2.

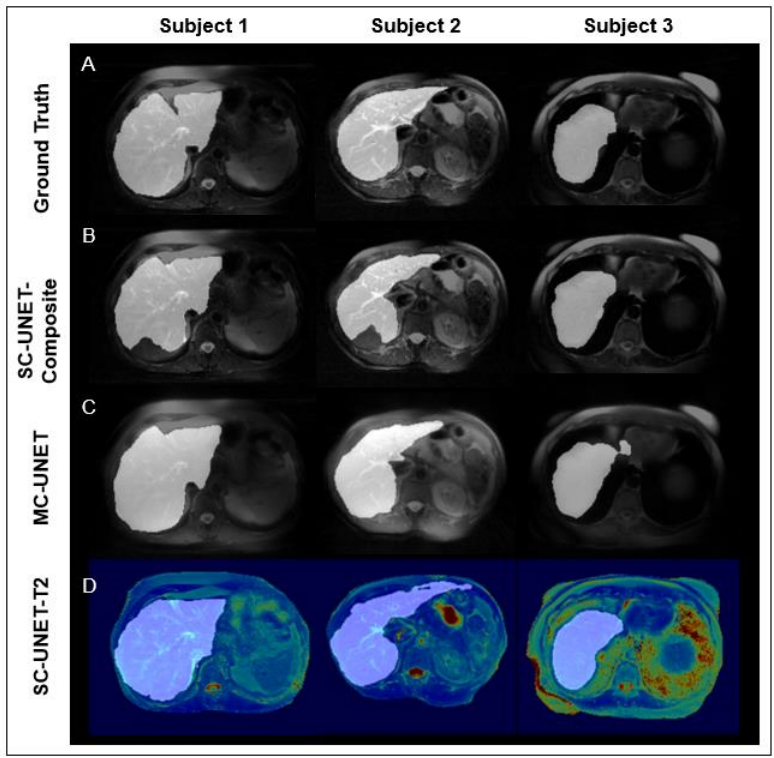

**Figure 4.** Comparison of image contrast between composite image (B), TE images (C and D), and T2 map (E) of a representative slice. The ground truth annotation for the liver is overlaid on the composite image in (A). The two TEs show images at a low echo time (~ 14ms) and a high echo time (~ 100ms). It can be observed that the T2-weighting in the image increases as we move from a lower TE to a higher one.